\documentclass{ifacconf}
\makeatletter
\let\old@ssect\@ssect 
\makeatother
\usepackage{amsmath}
\usepackage{amssymb}
\usepackage{enumerate}
\usepackage{definice}
\usepackage{color}
\usepackage{diagbox,multirow}
\usepackage{multirow}
\usepackage{float}

\usepackage[para]{threeparttable}
\usepackage{booktabs}
\usepackage{todonotes}

\usepackage{url}
\usepackage{hyperref}
\makeatletter
\def\@ssect#1#2#3#4#5#6{%
  \NR@gettitle{#6}
  \old@ssect{#1}{#2}{#3}{#4}{#5}{#6}
}
\makeatother
\usepackage{xspace}
\newcommand{\MATLAB}{\textsc{Matlab}\xspace}

\usepackage[scr=boondoxo,scrscaled=1.05]{mathalfa}

\def\p{^{L^+}}

\def\vv{^{(\operatorname{v})}}
\def\w{^{(\operatorname{w})}}

\def\l{^{L}}
\def\lm{^{L-1}}
\def\n{^{N}}
\def\ln{^{L,N}}
\def\pm{^{P-1}}
\def\p{^{P}}

\def\rmU{\mathrm U}

\usepackage{graphicx}      
\usepackage{natbib}        

\newcommand{\tightoverset}[2]{%
  \mathop{#2}\limits^{\vbox to -.5ex{\kern-0.75ex\hbox{$#1$}\vss}}}
  
\begin{document}
\begin{frontmatter}

\title{\hspace{-0mm}Noise Covariances Identification by MDM: Weighting, Recursion, and Implementation}
\author[First]{Oliver Kost, Jind\v{r}ich Dun\'{i}k, Ond\v{r}ej Straka} 

\address[First]{Department of Cybernetics and Research Centre NTIS, \\Faculty of Applied
Sciences, University of West Bohemia, \\\hspace*{-1.4cm}\mbox{Univerzitn\'{i} 8, 306 14 Pilsen, Czech Republic (e-mail: \{kost,dunikj,straka30\}@kky.zcu.cz)}}

\begin{abstract} 
The problem of noise covariance matrix identification of stochastic linear time-varying state-space models is addressed. 
The measurement difference method (MDM) is generalized to time-varying dimensions of the measurement and control.
Three MDM identification techniques that differ in weighting used in the underlying least squares method are proposed.
The techniques differ in estimate quality and computational complexity.
In addition, recursive forms are designed for two techniques.
The performance of the proposed techniques is analyzed using two numerical examples.
The implementation of techniques is enclosed with the paper.\footnote{\copyright 2024 the authors. This work has been accepted to IFAC for publication under a Creative Commons	Licence CC-BY-NC-ND}
\end{abstract}


\end{frontmatter}

\section{Introduction}

The availability of a quality model of a real system is a prerequisite for the proper functioning of an optimal design of modern control, fault-detection, and signal-processing
algorithms. 
Poor system modeling often leads to insufficient quality of the results or even algorithm failure.

The stochastic state-space model describes both deterministic and stochastic phenomena governing the behavior of the system quantities (called state) and the relation between the state and the measurement produced by sensors.
Hence, the stochastic state-space model model can be seen as a composition of two submodels.
The first one expresses deterministic system behavior arising often from the first principles, based on physical, kinematical, chemical, mathematical, biological, or other laws and rules.
The second one expresses the uncertain behavior of the state and measurement, where the uncertainty is modeled by noises described in the stochastic framework.
The uncertain effect is difficult to model using the first principles, and instead, it has to be identified from data.

Therefore, since the seventies, intensive research has focused on the design of methods identifying properties of the state and measurement noises.
In the literature, four classes of noise identification methods are recognized \citep{Meh:72}, \citep{DuStKoHa:17}, namely, Bayesian \cite{SaNu:09}, \citep{HuZyShCh:21}, covariance matching \citep{MoKi:93} \citep{SoHaIlAbBi:14}, maximum likelihood \citep{ScWiNi:11}, and correlation methods.
Compared with other classes, the correlation methods are often derived analytically with minimal additional assumptions related to the model and noise distribution.
This article focuses on the correlation methods, namely, on the measurement difference method (MDM) recently introduced in a series of articles 
\citep{KoDuSt:17}, \citep{DuKoSt:18}, \citep{KoDuSt:18}, \citep{KoDuSt:18a}, \citep{KoDuSt:18b}, \citep{DuKoStBl:20}, \citep{KoDuSt:21}, \citep{KoDuSt:22}, \citep{KoDuStDa:23}.
The MDM has been designed to identify the complete noise description for linear and nonlinear models with white or time/mutually correlated state and measurement noises. The MDM has been designed for problems with fixed measurement and vector and control vector dimensions. There are, however, problems where the dimensions may vary in time. Such problems often arise in navigation due to, e.g., the varying sensor availability or the varying visibility of satellites in GNSS positioning.

The paper proposes a generalization of the MDM for the problems with the time-varying dimensions of measurement and control. For these problems, three MDM identification techniques that differ in the weighting in the associated least squares method are proposed. In addition, the recursive forms of two identification techniques are proposed. The article is enclosed with the implementation of the techniques in both non-recursive and recursive forms.


The paper is structured as follows: Section~\ref{sec:noise_covariance_identification}
defines the covariance matrix identification problem. Section~\ref{sec:mdm} introduces the measurement difference method. Section~\ref{sec:mdm_identification_techniques} describes the proposed MDM identification techniques in both non-recursive and recursive versions. Section~\ref{sec:source} provides details about the implementation; Section~\ref{sec:numerical_illustration} gives the performance evaluation of the proposed MDM identification techniques, and Section~\ref{sec:conclusion} concludes the paper.
\section{Noise Covariance Matrix Identification Problem}  \label{sec:noise_covariance_identification}
Consider the following linear time-varying (LTV) discrete-in-time stochastic dynamic model in state-space representation with additive noises
\begin{subequations}\label{model}
\begin{align}
\bfx_{k+1}&=\bfF_k\bfx_k+\bfG_k\bfu_k+\bfE_k\bfw_k,\label{eqSt}\\
\bfz_{k}&=\bfH_k\bfx_k+\bfD_k\bfv_k,\label{eqMe}
\end{align}
\end{subequations}
for $ k=0,1,2,\ldots,\tau $. The vectors $\bfx_k\in\real^{n_x}$, $\bfu_k\in\real^{n_{u_k}}$, and $\bfz_k\in\real^{n_{z_k}}$ represent the immeasurable state of the system, the available control, and the available measurement at time instant $k$, respectively. The state $\bfF_k\in\real^{n_x\times n_x}$, control $\bfG_k\in\real^{n_x\times n_{u_k}}$, state-noise-shape $\bfE_k\in\real^{n_x\times n_w}$, measurement $\bfH_k\in\real^{n_{z_k}\times n_x}$, and measurement-noise-shape $\bfD_k\in\real^{n_{z_k}\times n_v}$ matrices are known and bounded. The system state is assumed to be \textit{observable}, i.e.,
\begin{align}
\rank\left(
\begin{bmatrix}
	\bfH_k \\ \bfH_{k+1}\bfF_k \\ \vdots \\ \bfH_{k+n_x-1}\underbrace{\bfF_{k+n_x-2}\!\hdots\bfF_k}_{{n_x-1}\text{ matrices}}
\end{bmatrix}
\right)=n_x\label{eq:obs},\ \forall k.
\end{align}
The state $\bfw_k\in\real^{n_w}$ and measurement $\bfv_k\in\real^{n_v}$ noises are assumed to be zero-mean\footnote{Estimation of non-zero mean of state and measurement noises is feasible and was solved in \cite{KoDuSt:18} and \cite{KoDuSt:22}} random variables with \textit{unknown} but bounded covariance matrices $\bfQ\in\real^{n_w\times n_w}$ and $\bfR\in\real^{n_v\times n_v}$.
In this paper, the problem with time-varying dimensions of the measurement and control input is considered, i.e., $n_{z_k}$ and $n_{u_k}$ may change in time.

The aim of noise matrix covariance identification is the estimation of covariance matrices $\bfQ$ and $\bfR$  from a set of measurements $\bfz_k$ and control inputs $\bfu_k$. 

\section{Measurement Difference Method}\label{sec:mdm}
The \textit{measurement difference method} is the correlation method for the noise properties identification \citep{DuKoSt:18}, \citep{KoDuSt:22}. 
Like any other correlation method, the MDM is based on analyzing the \textit{residual} (or measurement prediction error) vector, which is a linear function of the state and measurement noise vectors. The linearity of the relation between available residual and the noise vectors, which description is sought, is a key finding that facilitates the identification of noise properties (e.g., the noise covariance matrix (CM)) using the least-squares (LS) method.

In this section, the MDM 
is briefly reviewed, and its form for the case of time-varying dimensions of the measurement and control input is proposed. 
The proposed form applies to various problems, e.g., in chemical processes or navigation systems \citep{KoDuStDa:23}.

\subsection{Residual definition}
Derivation of the MDM starts with a definition of the augmented measurement vector as
\begin{multline}
	\bfZ_k\l = \calO_{k}\l\bfx_k + \bfGamma_k\lm\calG_k\lm\bfU_{k}\lm \\+ \bfGamma_k\lm\scrE_k\lm\bfW_{k}\lm + \calD_k\l\bfV_k\l \label{Z}
\end{multline}
for $k=0,\ldots,\tau-L+1$ with $L\geq1$, where $\bfZ_{k}\l\in\real^{n_{z_k\l}}$,  $\calO_k\l\in\real^{n_{z_k\l}\times n_x}$, $\bfGamma_k\lm\in\real^{n_{z_k\l}\times (L-1)n_x}$, $\calG_k\lm\in\real^{(L-1)n_x\times n_{u_k\lm}}$, $\scrE_k\lm\in\real^{(L-1)n_x\times (L-1)n_w}$, and $\calD_k\l\in\real^{n_{z_k\l}\times Pn_v }$ are defined\footnote{Note that for $L=1$ the matrices $ \bfGamma_k^0$, $\calG_ {k}^0$, $ \bfU_{k}^0$, $ \scrE_k^0 $, and $ \bfW_ {k}^0$ do not appear in \eqref{Z} as it reduces to meas. equation \eqref{eqMe} i.e., $\bfZ_{k}^1=\bfz_k$.} as 
\begin{subequations}
\begin{align}
	\bfZ_{k}\l&\triangleq\left[\begin{smallmatrix} \bfz_k \\ \bfz_{k+1} \\[-4px] \vdots \\ \bfz_{k+L-1}\end{smallmatrix}\right],\ 
	\calO_k\l\triangleq\left[\begin{smallmatrix}
		\bfH_k \\ \bfH_{k+1}\bfF_{k} \\ \bfH_{k+2}\bfF_{k+1}\bfF_{k} \\[-4px] \vdots \\ \bfH_{k+L-1}\calF_{k}\lm
		\hspace{-2px}
	\end{smallmatrix}\right]\!,
\end{align}
\begin{align}
	\bfGamma_k\lm&\triangleq\left[\begin{smallmatrix}\bfnul_{n_{z_k}\times n_x} & \bfnul_{n_{z_k}\times n_x}& \cdots & \bfnul_{n_{z_k}\times n_x} 
		\\ \bfH_{k+1} & \bfnul_{n_{z_{k+1}}\times n_x} & \cdots & \bfnul_{n_{z_{k+1}}\times n_x} 
		\\ \bfH_{k+2}\bfF_{k+1} & \bfH_{k+2} & \cdots & \bfnul_{n_{z_{k+2}}\times n_x}
		\\[-4px] \vdots & \vdots & \hspace{-3px}\ddots & \vdots \\
		\\[-4px] \bfH_{k+L-1}\calF_{k+1}^{L-2} & \bfH_{k+L-1}\calF_{k+2}^{L-3} & \hspace{-0px}\cdots\hspace{-0px}& \bfH_{k+L-1} \end{smallmatrix}\right]\!,\!\!\!
\end{align}
\vspace{-2mm}
\begin{align}
	\calG_k\lm&\triangleq\left[\begin{smallmatrix}
		\\ \bfG_{k} & \bfnul_{n_x\times n_{u_k}} & \cdots &\bfnul_{n_x\times n_{u_k}}
		\\ \bfnul_{n_x\times n_{u_k}} & \bfG_{k+1} & \cdots & \bfnul_{n_x\times n_{u_k}}
		\\[-4px] \vdots & \vdots & \hspace{-3px}\ddots & \vdots \\
		\\[-4px] \bfnul_{n_x\times n_{u_k}} & \bfnul_{n_x\times n_{u_k}} & \hspace{-0px}\cdots\hspace{-0px}& \bfG_{k+L-2} \end{smallmatrix}\right]\!,
\end{align}
\vspace{-2mm}
\begin{align}
	\scrE_k\lm&\triangleq\left[\begin{smallmatrix}
		\\ \bfE_{k} & \bfnul_{n_x\times n_w} & \cdots &\bfnul_{n_x\times n_w}
		\\ \bfnul_{n_x\times n_w} & \bfE_{k+1} & \cdots & \bfnul_{n_x\times n_w}
		\\[-4px] \vdots & \vdots & \hspace{-3px}\ddots & \vdots \\
		\\[-4px] \bfnul_{n_x\times n_w} & \bfnul_{n_x\times n_w} & \hspace{-0px}\cdots\hspace{-0px}& \bfE_{k+L-2} \end{smallmatrix}\right]\!,
\end{align}
\vspace{-2mm}
\begin{align}
	\calD_k\l&\triangleq\left[\begin{smallmatrix}
		\\ \bfD_{k} & \bfnul_{n_{z_{k}}\times n_v} & \cdots &\bfnul_{n_{z_{k}}\times n_v}
		\\ \bfnul_{n_{z_{k+1}}\times n_v} & \bfD_{k+1} & \cdots & \bfnul_{n_{z_{k+1}}\times n_v}
		\\[-4px] \vdots & \vdots & \hspace{-3px}\ddots & \vdots \\
		\\[-4px] \bfnul_{n_{z_{k+L-1}}\times n_v} & \bfnul_{n_{z_{k+L-1}}\times n_v} & \hspace{-0px}\cdots\hspace{-0px}& \bfD_{k+L-1} \end{smallmatrix}\right]\!,     
\end{align}
\end{subequations}
where $\calF_{k}^{j}\triangleq\prod_{i=1}^j\hspace{-2px}\bfF_{k+j-i}=\bfF_{k+j-1}\hdots\bfF_{k+1}\bfF_{k}\in\real^{n_x\times n_x}$, $ n_{z_k^L} = \Sigma_{i=0}^{L-1}n_{z_{k+i}}$, $ n_{u_k^{L-1}} = \Sigma_{i=0}^{L-2}n_{u_{k+i}}$, and $\bfnul_{n_i\times n_j}$ denotes a zero matrix of dimension $n_i\times n_j$. The symbol $\bfZ_k^L$, thus, represents a sequence of $L$ measurements $\bfz_k, \bfz_{k+1},\ldots,\bfz_{k+L-1}$ stacked columnwise. The vectors $\bfW_{k}\lm\in\real^{(L-1)n_w}$, $\bfV_{k}\l\in\real^{Ln_v}$, and $\bfU_{k}\lm\in\real^{n_{u_k\lm}}$ are defined analogously to $\bfZ_{k}\l$.

The MDM residuum $\tbfZ_k\in\real^{n_{z_k\l}}$ is defined the difference of the augmented measurement vector $\bfZ_k\l$ and its $ N $-step prediction as
\begin{align}
	\tbfZ_k
	&\triangleq\lefteqn{\bfZ_k\l-\hspace{-1mm}\overbrace{\phantom{\calH_{k-N}\ln\bfZ\l_{k-N}-\calA_{k}\w\calG_{k-N}\pm \bfU_{k-N}\pm}}^{N\text{-step predictive estimate of } \bfZ_k\l}}
	\underbrace{\phantom{\bfZ_k\l-}\calH_{k-N}\ln\bfZ\l_{k-N}}_{\calA_{k}\vv\bfZ\p_{k-N}}-\calA_{k}\w\calG_{k-N}\pm \bfU_{k-N}\pm\label{Zmdm}
\end{align}
where $\calA_{k}\w\in\real^{n_{z_k^L}\times (P-1)n_x}$ and ${\calA_{k}\vv}\in\real^{n_{z_k^L}\times n_{z_{k-N}^P}}$ being
\begin{subequations}\label{eq:Awv}
\begin{align}
	\calA_{k}\w&\triangleq\begin{bmatrix}\bfI_{n_{z_k^L}}, \bfI_{n_{z_{k}^L}}\end{bmatrix}\!\!
	\begin{bmatrix}
		\\[-9px]
		\begin{bmatrix}
			\calO_k\l\bfPhi_{k-N+1}\n,\bfGamma_k\lm
		\end{bmatrix}
		\\[0.5em]
		\!\begin{bmatrix}
			-\calH_{k-N}\ln\!\bfGamma_{k-N}\lm,
			\bfnul_{n_{z_{k}\l}\times N n_x}
		\end{bmatrix}\!\\[6px]\end{bmatrix},\label{eq:Aw}\\
	{\calA_{k}\vv}&\triangleq\begin{bmatrix}\bfI_{n_{z_k^L}}, \bfI_{n_{z_{k}^L}}\end{bmatrix}\!\!
	\begin{bmatrix}
		\begin{bmatrix}
			\bfnul_{n_{z_{k}^L}\times n_{z_{k-N}\n}}, \bfI_{n_{z_{k}^L}}
		\end{bmatrix}
		\\[0.5em]
		\!\begin{bmatrix}
			-\calH_{k-N}\ln,\bfnul_{n_{z_{k}^L}\times n_{z_{k-N+L}\n}}
		\end{bmatrix}\!\end{bmatrix}\label{eq:Av}.
\end{align}
\end{subequations}
where $ \calH_{k-N}\ln = \calO_k\l\calF_{k-N}^{N}\left(\calO_{k-N}\l\right)^\dagger\in\real^{n_{z_k\l}\times n_{z_{k-N}\l}},\\
 \bfPhi_{k-N+1}\n=[\calF_{k-N+1}^{N-1},\calF_{k-N+2}^{N-2},\hdots,\calF_{k-1}^{1},\bfI_{n_x}]\vspace{3px}\!\in\!\real^{n_x\times Nn_x}$, $\bfI_{n_x}$ denotes the identity matrix of dimension $n_x\times n_x$, and $N\geq0$\footnote{Note that for $N=0$ the matrices $ \bfPhi_{k-N+1}\n$ and $\calF_{k-N}^{N} $ do not appear in equations \eqref{Zmdm} and \eqref{eq:Awv} i.e., 
 	$ \calA_{k}\vv=\bfI_{n_{z_{k}^L}}\!-\calO_k\l\left(\calO_{k}\l\right)^\dagger,$
 	$\calA_{k}\w\!\!=\!\calA_{k}\vv\bfGamma_k\lm\!,\!$ 
	$\tbfZ_k\!=\!\calA_{k}\vv\bfZ\l_{k}\!-\calA_{k}\w\calG_{k}\lm \bfU_{k}\lm\!\!=\!\calA_{k}\w\scrE_{k}\lm \bfW_{k}\lm \!\ \linebreak+\calA_{k}\vv\calD_{k}\l \bfV_{k}\l $.
}. 
The value $L\geq1$ is a user-defined parameter chosen sufficiently large to ensure the full column rank of the matrix $\calO_k\l$ and $P=L+N$. 

The term $ \calA_{k}\vv\bfZ_{k-N}\p $ in \eqref{Zmdm} can be rewritten as \citep{KoDuSt:22}
\begin{multline}
	\calA_{k}\vv\bfZ_{k-N}\p={\calA_{k}\w}\scrE_{k-N}\pm \bfW_{k-N}\pm+{\calA_{k}\vv}\calD_{k-N}\p \bfV_{k-N}\p\\+{\calA_{k}\w}\calG_{k-N}\pm \bfU_{k-N}\pm.
\end{multline}
This allows the residuum $\tbfZ_k$ \eqref{Zmdm} to be written in a compact form as a function of the sequence of available measurements and controls but also as a function of the sequence of unknown state and measurement noises as
\begin{subequations}
\begin{align}
	\!\!\!\tbfZ_k&=\calA_{k}\calB_{k}\begin{bmatrix}-\bfU_{k-N}\pm\\\hspace{3mm}\bfZ_{k-N}\p\end{bmatrix}\!,\!\!\label{mpeMDMZU}
	\\
	\!&=\calA_{k}\calC_{k}\begin{bmatrix}\bfW_{k-N}\pm \\ \bfV_{k-N}\p \end{bmatrix}=\calA_{k}\calC_{k}\calE_{k},\label{mpeMDMWV}
\end{align}
\end{subequations}
where the matrices $\calA_{k}\in\real^{n_{z_k^L}\times ((P-1)n_x+n_{z_{k-N}^P})}$,\linebreak $ \calB_{k}\in\real^{((P-1)n_x + n_{z_{k-N}\p})\times (n_{u_{k-N}\pm} + n_{z_{k-N}\p} )}$,\\ $ \calC_{k}\in\real^{((P-1)n_x + n_{z_{k-N}}\p )\times ( (P-1)n_w + Pn_v)}$, and the vector $\calE_{k}\in\real^{(P-1)n_w+Pn_v}$ are defined as
\begin{subequations}
\begin{align}
	\calA_{k}&\triangleq\begin{bmatrix}\calA_{k}\w & \calA_{k}\vv\end{bmatrix}, \ \ \ \ \ \calE_{k}\triangleq\begin{bmatrix}\bfW_{k-N}\pm \\ \bfV_{k-N}\p \end{bmatrix},
\\
\calB_{k} &\triangleq \begin{bmatrix}\calG_{k-N}\pm & \bfnul_{(P-1)n_x \times n_{z_{k-N}^P}}\!\!\\ \bfnul_{n_{z_{k-N}^P} \times n_{u_{k-N}\pm}} & \bfI_{n_{z_{k-N}^P}}\end{bmatrix},
\\
\calC_{k} &\triangleq \begin{bmatrix}\scrE_{k-N}\pm & \bfnul_{(P-1)n_x \times Pn_v}\!\\ \bfnul_{n_{z_{k-N}\p} \times (P-1)n_w} & \calD_{k-N}\p \end{bmatrix}.
\end{align}
\end{subequations}

\subsection{Covariance of residuum and noises}
The available residuum $\tbfZ_k$ \eqref{mpeMDMZU} is a \textit{linear} function of the unavailable state and measurement noises as was shown in \eqref{mpeMDMWV}. Consequently, the \textit{time-varying} covariance of the residuum for the time instant $k$ is a linear function of the \textit{time-invariant} covariances of the state and measurement noises, i.e.,
\begin{align}
	\calR_{\tbfZ_k^2}&=\mean\left[\tbfZ_k^{\otimes^2}\right]=
	\calA_{k}^{\otimes^2}\calC_{k}^{\otimes^2}
	\calR_{\calE^2},\label{noncentMom}
\end{align}
where $\otimes$ denotes the Kronecker product, $\bfB^{\otimes^n}\!\!\triangleq\!\overbrace{\bfB\!\otimes\!\ldots\otimes\!\bfB}^{n-\text{terms}}$ denotes the Kronecker power, $\calR_{\tbfZ_k^2}\in\real^{(n_{z_k\l})^2}$ is residual \textit{covariance} (vector) containing the CM elements of the residuum $\tbfZ_k$ and $\calR_{\calE^2}\in\real^{((P-1)n_w+Pn_v)^2}$ is noise covariance (vector) containing CM elements of the augmented noise vector $ \calE_{k} $, respectively, state and measurement noises.

The noise covariance $\calR_{\calE^2}$ contains multiple copies of \textit{unique} elements of the noise CMs gathered in the vector denoted as $\calR_{\calE^2}^\rmU$, which are related via the replication matrix $\Psi_{\calE^2}$ as
\begin{align}
	\calR_{\calE^2}&=\Psi_{\calE^2}\calR_{\calE^2}^\rmU,\label{NpsiM}
\end{align}

The residue covariance $\tbfZ_k$ contains redundant copies of the same elements. Therefore, unique elements of the vector $\calR_{\tbfZ_k^2}$ are selected using a unification matrix \linebreak $\Xi_{\tbfZ_k^2}\in\real^{(n_{z_k\l}(n_{z_k\l}+1)/ 2)\times(n_{z_k\l})^2}$ as  
\begin{align}
	\calR_{\tbfZ_k^2}^{\rmU}=\Xi_{\tbfZ_k^2}\calR_{\tbfZ_k^2}\label{NXiM}
\end{align}

The equations \eqref{NpsiM} and \eqref{NXiM} can be used to modify \eqref{noncentMom} into the following form
\begin{align}
\calR_{\tbfZ_k^2}^{\rmU}&=\Xi_{\tbfZ_k^2}\calA_{k}^{\otimes^2}\calC_{k}^{\otimes^2}\Psi_{\calE^2}\calR_{\calE^2}^\rmU.\label{NmomMAllk}
\end{align}

\subsection{Covariance of residuum estimation}
The covariance of the residue $ \calR_{\tbfZ_k^2}^{\rmU} $ is unknown.
However, it is possible to compute its estimate $ \tbfZ_k^{\otimes^2} $ and the relation \eqref{NmomMAllk} can be modified as
\begin{align}
	\underbrace{\Xi_{\tbfZ_k^2}\tbfZ_k^{\otimes^2}}_{\hspace{-20px}\text{Known vector}\hspace{-20px}}
	=
	\underbrace{\Xi_{\tbfZ_k^2}\calA_{k}^{\otimes^2}\calC_{k}^{\otimes^2}\Psi_{\calE^2}}_{\text{Known matrix}}
	\overbrace{\calR_{\calE_k^2}^\rmU}^{\hspace{-50px}\text{Sought noise covariances}\hspace{-50px}}
	+
	\underbrace{\Xi_{\tbfZ_k^2}\calA_{k}^{\otimes^2}\calC_{k}^{\otimes^2}\!}_{\text{Known matrix}}\
	\overbrace{\bfeta_k}^{\hspace{-20px}\text{Unknown vector}\hspace{-20px}},\hspace{-5px}\label{eq:wls2}
\end{align}
where $\bfeta_k=\calE_k^{\otimes^2}-\Psi_{\calE^2}\calR_{\calE^2}^\rmU$ is the \textit{zero-mean} process with the covariance and ``cross-covariance'' matrices in the vector form defined as
\begin{subequations}\label{eq:wls3}
	\begin{align}
		\calR_{\bfeta^2}&=\mean\left[\bfeta_{k}^{\otimes^2}\right]=\mean\left[\calE_{k}^{\otimes^4}\right]-\calR_{\calE^2}^{\otimes^2},\label{eq:wls3a}\\
		\calR_{\bfeta_{k},\bfeta_{j}}&=\mean\left[\bfeta_{k}\otimes\bfeta_{j}\right]=\mean\left[\calE_{k}^{\otimes^2}\otimes\calE_{j}^{\otimes^2}\right]
		-\calR_{\calE^2}^{\otimes^2}.\label{eq:wls3b}
	\end{align}
\end{subequations}
The equation \eqref{eq:wls2} can be written for all possible time instants $k=N,\ldots,\tau-L+1$ as a function of unique elements of the noise covariance \eqref{NpsiM} in a compact form as
\begin{align}
	\mathscr{R}_{\tbfZ^2}&=\mathscr{A}_{2}\calR_{\calE^2}^\rmU + \scrL_2\scrY_2,\label{NmomMAll}
\end{align}
where
\begin{subequations}
\begin{align}
	\!\scrR_{\tbfZ^2}\!\!=\!\!\!
	\begin{bmatrix}
			\Xi_{\tbfZ_N^2}\tbfZ_N^{\otimes^2}\\
			\Xi_{\tbfZ_{N+1}^2}\tbfZ_{N+1}^{\otimes^2}\\
			\vdots\\
			\!\Xi_{\tbfZ_{\tau-L+1}^2}\tbfZ_{\tau-L+1}^{\otimes^2}\!\!\!
			\\[2mm]
		\end{bmatrix}\!\!\!,
	\scrA_{2}\!=\!\!\!
	\begin{bmatrix}
		\Xi_{\tbfZ_N^2}\calA_{N}^{\otimes^2}\calC_{N}^{\otimes^2}\\
		\Xi_{\tbfZ_{N+1}^2}\calA_{N+1}^{\otimes^2}\calC_{N+1}^{\otimes^2}\\
		\vdots\\
		\!\Xi_{\tbfZ_{\tau-L+1}^2}\!\calA_{\tau-L+1}^{\otimes^2}\calC_{\tau-L+1}^{\otimes^2}\!\!
	\end{bmatrix}\!\!\!\Psi_{\calE^2},
\end{align}
\vspace{-2mm}
\begin{align}
	\scrL_2 &=
	\begin{bmatrix}
			\Xi_{\tbfZ_{N}^2}\calA_{N}^{\otimes^2}\calC_{N}^{\otimes^2} & \hdots & \bfnul\\
			\vdots & \ddots & \vdots \\
			\bfnul & \hdots & \Xi_{\tbfZ_{\tau-L+1}^2}\calA_{\tau-L+1}^{\otimes^2}\calC_{\tau-L+1}^{\otimes^2}
		\end{bmatrix}
	\\
	\scrY_2 &=
	\begin{bmatrix}
			\bfeta_N\T & \bfeta_{N+1}\T &
			\hdots & 
			\bfeta_{\tau-L+1}\T
	\end{bmatrix}\T
	\!\!\label{Y2}
\end{align}
\end{subequations}
and $\scrR_{\tbfZ^2}$ is a sample-based autocovariance function of $\tbfZ_k$.
\section{MDM Implementations}\label{sec:mdm_identification_techniques}
In the previous section, \eqref{NmomMAll} relates the noise covariance and residue covariance estimate through a linear equation involving known model matrices.
Several techniques based on the LS method can be used to estimate the noise covariance $ \calR_{\calE^2}^\rmU $.
In this section, the following three MDM identification techniques are proposed:
\begin{enumerate}
	\item unweighted,
	\item weighted,
	\item semi-weighted.
\end{enumerate}
In addition, recursive forms of the unweighted and semi-weighted MDM identification techniques are also given.

\subsection{Unweighted identification}
The MDM unweighted identification (non-recursive) (Uw-Nr) estimates the noise covariances, assuming known full-rank matrix $\scrA_2$, using \eqref{NmomMAll} as
\begin{align}
    \widehat{\left(\calR_{\calE^2}^\rmU\right)^\text{Uw-Nr}}=\left(\scrA_2\T\scrA_2\right)^{\!-1}\!\!\scrA_2\T{\scrR_{\tbfZ^2}},\label{odhadEmN}
\end{align}

\subsection{Weighted identification with CM of the estimate}
The MDM weighted non-recursive least-squares (We-Nr) identification is based on the weighted LS method. 
For weighted estimation of noise covariance using \eqref{NmomMAll}, it is necessary to know the CM of $ \scrL_2\scrY_2 $, which is given by
\begin{align}
	\mathcal{P}_2
	&=
	\scrL_2\!
	\underbrace{\!\begin{bmatrix}
		(\calR_{\bfeta^2})_\mathsf{M} & \ldots & (\calR_{\bfeta_{\tau-L+1},{\bfeta_N}})_\mathsf{M}\\
		\vdots & \ddots & \vdots\\
		(\calR_{\bfeta_{N},\bfeta_{\tau-L+1}})_\mathsf{M} & \ldots & (\calR_{\bfeta^2})_\mathsf{M}\\
	\end{bmatrix}\!}
	_{\mean\left[\scrY_2\scrY_2\T\right]}
	\!\scrL_2\T,\label{eq:wls5}
\end{align}
where the notation $(\cdot)_\mathsf{M}$ stands for the inverse of the column-wise stacking of the matrix denoted by $(\cdot)_\mathsf{V}$, i.e., it holds $(\bfB_\mathsf{V})_\mathsf{M}=\bfB$. From \eqref{eq:wls3} and \eqref{eq:wls5} can be seen that CM $ \mathcal{P}_2 $ is a function of not only the sought noise covariances $ \calR_{\calE^2}$ but also of the fourth moments of the noises $ \mean\left[\calE_{k}^{\otimes^4}\right]$ and $ \mean\left[\calE_{k}^{\otimes^2}\otimes\calE_{j}^{\otimes^2}\right] $.
Unfortunately, these moments are unknown. 
Therefore, the following sequential technique is proposed:
\begin{itemize}
	\item Compute the noise covariance estimates by the standard MDM identification, e.g., Uw-Nr MDM estimate using \eqref{odhadEmN} or, as will be shown in the following section, a semi-weighted MDM estimate.
	\item If the noises are Gaussian\footnote{For the Gaussian noises, it is sufficient to estimate only the noise covariance $ \calR_{\calE^2}$ and the fourth noise moments $ \mean\big[\!\calE_{k}^{\otimes^4}\big]\!, \mean\big[\!\calE_{k}^{\otimes^2}\!\!\otimes\!\calE_{j}^{\otimes^2}\big] $\linebreak can be computed on their basis. For other distributions, one can estimate both the noise covariance and the fourth moments using MDM \citep{KoDuSt:22}.}, the estimated noise covariances can be used for calculation of the approximate noise moment estimates \eqref{eq:wls3}, further denoted as $\widehat{\calR_{\bfeta^2}}$, $\widehat{\calR_{\bfeta_{k},\bfeta_{j}}}$.
	\item  Substitution of ${\calR_{\bfeta^2}}$, ${\calR_{\bfeta_{k},\bfeta_{j}}}$ by the approximate noise moments $\widehat{\calR_{\bfeta^2}}$, $\widehat{\calR_{\bfeta_{k},\bfeta_{j}}}$ according to \eqref{eq:wls5} with $\widehat{\mathcal{P}_2}$ in place of $\mathcal{P}_2$. 
\end{itemize}
Then, the MDM We-Nr identification is given by
\begin{align}
	\widehat{\left(\calR_{\calE^2}^\rmU\right)^\text{We-Nr}}=\left(\scrA_2\T\widehat{\mathcal{P}_2}^{-1}\scrA_2\right)^{\!-1}\!\scrA_2\T\widehat{\mathcal{P}_2}^{-1}{\scrR_{\tbfZ^2}}.\label{eq:wls6}
\end{align}
Having the estimate of the CM $\widehat{\mathcal{P}_2}$, an approximate\footnote{The true and estimated CM of the noise covariance estimates are very similar i.e., $ \big(\!\scrA_2\T{\mathcal{P}_2}^{\!-1}\scrA_2\!\big)^{\!\!-1}\!\!\approx\!\!\big(\!\scrA_2\T\big(\widehat{\mathcal{P}_2}\big)^{\!\!-1}\!\scrA_2\!\big)^{\!\!-1} $.} CM of the MDM We-Nr estimate can be computed as 
\begin{align}
	\cov\left[\widehat{\left(\calR_{\calE^2}^\rmU\right)^\text{We-Nr}}\right]
	\approx
	\bigg(\scrA_2\T\left(\widehat{\mathcal{P}_2}\right)^{\!-1}\scrA_2\bigg)^{\!\!-1}.\label{eq:wls7}
\end{align}
\subsection{Semi-weighted identification}
The MDM semi-weighted non-recursive least-squares (Sw-Nr) identification is based on the MDM We-Nr identification however, the CM $ \mathcal{P}_2 $ is computed from \eqref{eq:wls5} where $ \mean\left[\scrY_2\scrY_2\T\right]=\bfI $ as
\begin{align}
	\mathcal{S}_2
	&=\scrL_2\scrL_2\T,
\end{align}
and the MDM Sw-Nr identification is given by
\begin{align}
	\widehat{\left(\calR_{\calE^2}^\rmU\right)^\text{Sw-Nr}}=\left(\scrA_2\T\mathcal{S}_2^{-1}\scrA_2\right)^{\!-1}\!\scrA_2\T\mathcal{S}_2^{-1}{\scrR_{\tbfZ^2}}.\label{semiEst}
\end{align}
Compared to the MDM We-Nr identification, the MDM Sw-Nr identification does not need an estimate of the noise moments \eqref{eq:wls3}; however, it typically provides an estimate with similar\footnote{The MDM We-Nr and Sw-Nr estimates are very similar, i.e., $ \widehat{\big(\calR_{\calE^2}^\rmU\big)^\text{We-Nr}}\approx \widehat{\big(\calR_{\calE^2}^\rmU\big)^\text{Sw-Nr}} $. In contrast, the estimated covariance matrices of the noise covariance estimates differ, i.e., $ \big(\!\scrA_2\T\big(\widehat{\mathcal{P}_2}\big)^{\!\!-1}\!\scrA_2\!\big)^{\!\!-1}\!\!\neq\!\!\big(\!\scrA_2\T\big({\mathcal{S}_2}\big)^{\!\!-1}\!\scrA_2\!\big)^{\!\!-1} $.} quality as MDM We-Nr.

\subsection{Recursive unweighted/semi-weighted identification}
The MDM Uw-Nr and Sw-Nr (i.e., non-recursive) identifications use the ordinary LS method, which is easy to convert them to the recursive LS method, further denoted as Uw-Re and Sw-Re, respectively.
The calculation itself can be done using the following generalized relations
\begin{subequations}
\begin{align}
	\bfK_k &= \bfSigma_{k-1} \bfA_k\T\left(\bfA_k \bfSigma_{k-1} \bfA_k\T + \bfOmega_k \right)^{-1}
	\\
	\widehat{\left(\calR_{\calE^2}^\rmU\right)}_k \!&=\! \widehat{\left(\calR_{\calE^2}^\rmU\right)}_{k-1} 
	\!+ 
	\bfK_k \! \left(\Xi_{\tbfZ_k^2}\tbfZ_k^{\otimes^2} - \bfA_k\widehat{\left(\calR_{\calE^2}^\rmU\right)}_{k-1} \right)\label{rek1}\!\!\!
	\\
	\bfSigma_k &= \left(\bfI - \bfK_k \bfA_k \right) \bfSigma_{k-1}\label{rek2}
\end{align}
where $ \bfA_k = \Xi_{\tbfZ_k^2}\calA_{k}^{\otimes^2}\calC_{k}^{\otimes^2}\Psi_{\calE^2} $, the vector$ \widehat{\left(\calR_{\calE^2}^\rmU\right)}_k $ is estimate at time step $ k $, and the matrix $ \bfOmega_k $ depending on the choice of the MDM Uw-Re or Sw-Re identification is defined as
\begin{align}
	\!\bfOmega_k &= \begin{cases}
		\bfI_{n_{z_k\l}(n_{z_k\l}+1)/ 2}\hspace{23mm} \text{, for Uw-Re}
		\\
		\Xi_{\tbfZ_k^2}\calA_{k}^{\otimes^2}\calC_{k}^{\otimes^2}\left(\Xi_{\tbfZ_k^2}\calA_{k}^{\otimes^2}\calC_{k}^{\otimes^2}\right)\T \text{, for Sw-Re.}
	\end{cases} \!\!\!\!\!\!
\end{align}
\end{subequations}
Since the equation error vector $ \Xi_{\tbfZ_k^2}\calA_{k}^{\otimes^2}\calC_{k}^{\otimes^2}\bfeta_k$ \eqref{eq:wls2} is time-correlated and the CM 	$ \mathcal{P}_2 $ \eqref{eq:wls5} cannot be estimated with sufficient quality for a small number of measurements (i.e., for the beginning of the recursive/online estimation), it is not possible to straightforwardly convert the MDM We-Nr identification to a recursive version.

\section{Available MDM source codes}\label{sec:source}
An important component of the article is the publication of \MATLAB source code that computes all the necessary known matrices and vectors used by the MDM in \eqref{eq:wls2} i.e., $ \Xi_{\tbfZ_k^2}\tbfZ_k^{\otimes^2} $, $ \Xi_{\tbfZ_k^2}\calA_{k}^{\otimes^2}\calC_{k}^{\otimes^2}\Psi_{\calE^2}$, and $ \Xi_{\tbfZ_k^2}\calA_{k}^{\otimes^2}\calC_{k}^{\otimes^2}$. 
In addition, the source code provides the variables $ \calR_{\bfeta^2}$  $\calR_{\bfeta_{\tau-L+1},{\bfeta_N}} $ in \eqref{eq:wls5} as functions of the (estimated) second \textit{Gaussian} noise covariances $ \calR_{\calE^2} $ \eqref{eq:wls3}  i.e., $\calR_{\bfeta_{k+j},{\bfeta_k}}=\bff_j(\calR_{\calE^2})$, for $ j=0,1,\hdots ,L+N-1$.

The code is available at:\\ \url{https://github.com/IDM-UWB/MDM_SYSID2024}

\section{Numerical Illustrations}\label{sec:numerical_illustration}
Two dynamic LTV models \eqref{model} are considered with the initial state being Gaussian with $\mean\left[x_0\right]=1$ and $\var\left[x_0\right]=1$, the control is $u_k=\sin(k/\tau)$, and the state and measurement independent noises are Gaussian. The MDM is analyzed using $10^4$ Monte-Carlo (MC) simulations with $\tau=10^3$ measurement samples per MC simulation.

\subsection{Comparison of MDM weighting techniques}
\begin{figure}[h]
	\hspace{-4mm}
	\includegraphics[scale=0.61]{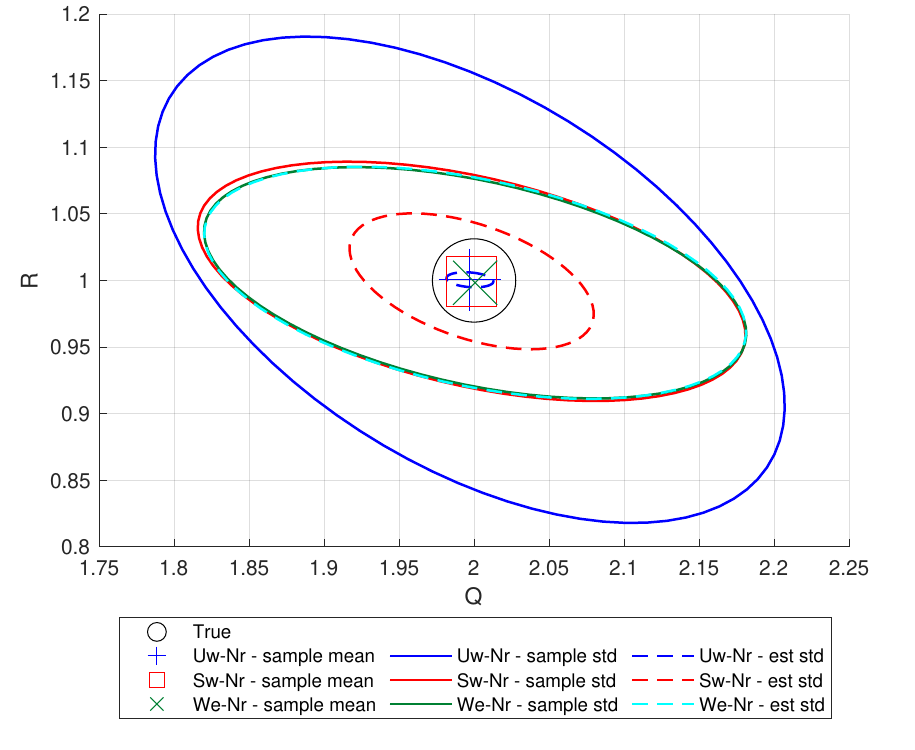}
	\vspace*{-9mm}
	\caption{MDM un/semi/weighted non-recursive ident.}
	\label{tacFig}
	\vspace*{-1mm}
\end{figure}

This section compares the non-recursive MDM unweighted, semi-weighted, and weighted identification techniques using a simulation example. The dynamic LTV model from \cite{KoDuSt:22}\!\footnote{The sign in the measurement equation is incorrect in \cite{KoDuSt:22}. Alternatively, the resulting estimate is changed negligibly.} is used and is given as follows
\begin{subequations}
\begin{align}
	F_k&=0.8-0.1\sin(7\pi k / \tau),\\
	H_k&=1+0.99\sin(100\pi k / \tau),\\
	G_k&=E_k=D_k=1,\ \ \ \ Q=2,\ \ \ \ R=1,\\
	n_x&=n_{u_k}=n_w=n_{z_k}=n_v=1,\ \ \ \ L=N=1 
\end{align}
\end{subequations}
The sample mean (S. mean) and the diagonal of the sample covariance matrix (S. cov) of MDM Uw-Nr, Sw-Nr, and We-Nr estimates per MC are shown in Table \ref{tableTAES}. 
In addition, the average computation time and the diagonal of the sample mean of the estimated covariance matrix (est. cov) of the noise covariance estimation are shown in Table \ref{tableTAES}, which is computed by 
\begin{itemize}
	\item $ \cov\left[\widehat{\big(\calR_{\calE^2}^\rmU\big)^\text{\!Uw-Nr}}\right] \approx \big(\scrA_2\T\scrA_2\big)^{\!\!-1} $,\hspace{13mm} for Uw-Nr,\\[0.5mm] 
	\item $\cov\left[\widehat{\big(\calR_{\calE^2}^\rmU\big)^\text{Sw-Nr}}\right] \approx \big(\scrA_2\T\big(\widehat{\ \mathcal{S}_2} \ \big)^{\!-1}\scrA_2\big)^{\!\!-1} $,\hspace{0.5mm} for Sw-Nr,\\[0mm]
	\item and according to \eqref{eq:wls7},\hspace{28.9mm} for We-Nr.
\end{itemize}
Note that in the MDM We-Nr calculation, MDM Uw-Nr estimates are used to obtain the weighting matrix estimate $ \widehat{\mathcal{P}_2} $ in \eqref{eq:wls6} and \eqref{eq:wls7}.
%
%
%
%
%
%
%
%
%
\begin{table}[H]
	\vspace{-2mm}
	\hspace{-1mm}
	\scalebox{0.817}{
	\begin{tabular}{cccccccccccccc}
		\toprule
		& &\!\!\!\!\!& \multicolumn{3}{c}{Uw-Nr} & \!\!\!\!\!\! & \multicolumn{3}{c}{\hspace{-4px}Sw-Nr} & \!\!\!\!\!\! & \multicolumn{3}{c}{\hspace{-4px}We-Nr(Uw-Nr)}\\
		& \hspace{-7px}True\hspace{-7px} & \!\!\!\!\!\! 
		& \hspace{-9px} S.\! mean & \hspace{-15px} S.\! cov \hspace{-15px} & \hspace{-10px} est.\! cov\hspace{-10px} & \!\!\!\!\!\! 
		& \hspace{-9px} S.\! mean & \hspace{-15px} S.\! cov \hspace{-15px} & \hspace{-10px} est.\! cov\hspace{-10px} & \!\!\!\!\!\! 
		& \hspace{-9px} S.\! mean & \hspace{-15px} S.\! cov \hspace{-15px} & \hspace{-10px} est.\! cov\hspace{-10px}
		\\[-11px]
		\\ \cmidrule{1-2}\cmidrule{4-6} \cmidrule{8-10} \cmidrule{12-14}  
		\\[-10px]
		$Q$ & \!2\!   &\!\!\!\!\!\! & 1.997  &  \hspace{-5px}0.044\hspace{-5px} & 0.000 
		              &\!\!\!\!\!\! & 1.998  &  \hspace{-5px}0.033\hspace{-5px} & 0.007 
		              &\!\!\!\!\!\! & 2.001  &  \hspace{-5px}0.033\hspace{-5px} & 0.033\\
		$R$ & \!1\!   &\!\!\!\!\!\! & 1.001  &  \hspace{-5px}0.033\hspace{-5px} & 0.000 
		              &\!\!\!\!\!\! & 0.999  &  \hspace{-5px}0.008\hspace{-5px} & 0.003 
		              &\!\!\!\!\!\! & 0.998  &  \hspace{-5px}0.008\hspace{-5px} & 0.008
		\\[-11px]
		\\ \cmidrule{1-2}\cmidrule{4-6} \cmidrule{8-10} \cmidrule{12-14}  
		\\[-10px]
		\!\!\!\!\text{Time}\!\!\!\! & - &\!\!\!\!\!\! &   \multicolumn{3}{c}{1$\times$Uw-Nr}   &\!\!\!\!\!\! &    \multicolumn{3}{c}{9.46$\times$Uw-Nr}   &\!\!\!\!\!\! &    \multicolumn{3}{c}{2130$\times$Uw-Nr} \\
		\toprule
	\end{tabular}
	}
	\caption{Performance of un/semi/weighted MDM identification.}
	\label{tableTAES}
	\vspace*{-6mm}
\end{table}
The results shown in Table \ref{tableTAES} are also depicted in Fig. \ref{tacFig}.

Table \ref{tableTAES} and Fig. \ref{tacFig} show that the MDM Sw-Nr and We-Nr provide similar estimates with significantly lower covariance than the MDM Uw-Nr estimate. The single MDM We-Nr approach additionally provides \textit{correct} covariance of the estimates at the cost of higher computational complexity.

\subsection{Performance analysis of recursive forms}
This section compares the non-recursive and recursive forms of MDM unweighted and semi-weighted identification techniques. The following dynamic LTV model 
\begin{subequations}
\begin{align}
	F_k&=1+0.1\sin(20\pi k /\tau),\ \ \ G_k=1,\ \ \ E_k=-1,\\ 
	Q&=3,\ \ \ \bfR=\left[\begin{smallmatrix}\hspace{2mm}2 &\ -1\\ -1 &\hspace{3.4mm}1\end{smallmatrix}\right],\\ 
	n_x&=n_w=n_{u_k}=1,\ \ \ n_v=2,\ \ \ L=2,\ \ \ N=1\\ 
	n_{z_k}\!&=\!1,\ \ H_k\!=\!1,\hspace{3.3mm}\ \ \bfD_k\!=\!\begin{bmatrix} 1& 0\end{bmatrix}\!,\ \ \text{if } k<\tau/3\\
	n_{z_k}\!&=\!1,\ \ H_k\!=\!1,\hspace{3.3mm}\ \ \bfD_k\!=\!\begin{bmatrix} 0& 1\end{bmatrix}\!,\ \ \text{if } \tau/3\leq k< 2\tau\!/3\!\!\\
	n_{z_k}\!&=\!2,\ \ \bfH_k\!=\left[\begin{smallmatrix}1\\1\end{smallmatrix}\right]\!,\ \ \bfD_k\!=\!\bfI_2,\hspace{17px} \text{otherwise}
\end{align}
\end{subequations}
is considered. The sought noise covariances $ \calR_{\calE^2}^\rmU=\begin{bmatrix}	Q& \bfR(1,1) & \bfR(1,2) & \bfR(2,2) \end{bmatrix}\T $ are at the first time instant of the recursive technique chosen as $ 	\widehat{\left(\calR_{\calE^2}^\rmU\right)}_{\!0} \!=\!\begin{bmatrix}	0.5& 0.5& 0& 0.5\end{bmatrix}\T$\eqref{rek1} and $\ \bfSigma_0 \!=\!10\hspace{1px}\bfI_4 $\eqref{rek2}.

This model represents the case where the availability of measurements/sensors/signals changes over time.
The simulation time interval is split into three equal parts.
In the first one, only the first sensor is available; in the second one, only the second sensor is available; and in the last one, both sensors are available.

\begin{figure}[h]
    \vspace*{-4mm}
	\includegraphics[scale=0.74]{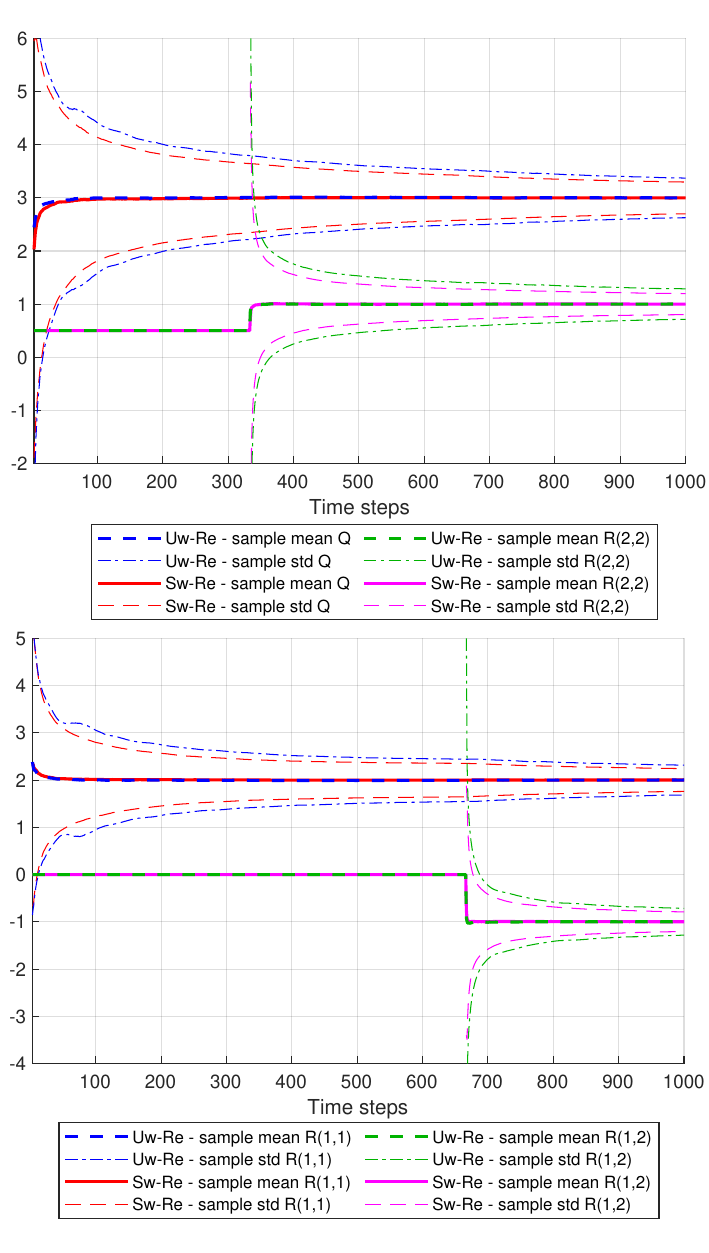}
	\vspace*{-10mm}
	\caption{Un/semi-weighted MDM non/recursive ident.}
	\label{recFigtime}
	\vspace*{-2mm}
\end{figure}

The S. mean and the diagonal of S. cov of MDM Uw-Nr, Uw-Re, Sw-Nr, and Sw-Re estimates per MC and the average computation time 
are shown in Table \ref{tableRec}
. The recursive estimates (i.e., MDM Uw-Re and Sw-Re) are shown for the last instant.  
%
%
%
%
%
%
%
%
\begin{table}[H]
    \vspace*{-1mm}
	\hspace{-0.5mm}
	\scalebox{0.83}{
		\begin{tabular}{cccccccccccccc}
			\toprule
			& &\!\!\!\!\!\!& \multicolumn{2}{c}{Uw-Nr} & \!\!\!\!\!\! & \multicolumn{2}{c}{\hspace{-4px}Uw-Re} &\!\!\!\!\!\!& \multicolumn{2}{c}{Sw-Nr} & \!\!\!\!\!\! & \multicolumn{2}{c}{\hspace{-4px}Sw-Re} \\
			& \hspace{-7px}True\hspace{-5px} & \!\!\!\!\! 
			& \hspace{-8px} S. mean & \hspace{-15px} S. cov \hspace{-15px} & \!\!\!\!\!\! 
			& \hspace{-8px} S. mean & \hspace{-15px} S. cov \hspace{-15px} & \!\!\!\!\!\! 
			& \hspace{-8px} S. mean & \hspace{-15px} S. cov \hspace{-15px} & \!\!\!\!\!\! 
			& \hspace{-8px} S. mean & \hspace{-30px} S. cov \hspace{-20px} \!\!\!\!\!\!
			\\[-11px]
			\\ \cmidrule{1-2} \cmidrule{4-5} \cmidrule{7-8} \cmidrule{10-11}  \cmidrule{13-14} 
			\\[-10px]
			\!\!\!$Q$\!\!\! & \!3\!        
			&\!\!\!\!\!\! &  \hspace{-5px}3.000\hspace{-5px}  &  \hspace{-5px}0.139\hspace{-5px} 
			&\!\!\!\!\!\! &  \hspace{-5px}2.999\hspace{-5px}  &  \hspace{-5px}0.139\hspace{-5px} 
			&\!\!\!\!\!\! &  \hspace{-5px}3.002\hspace{-5px}  &  \hspace{-5px}0.090\hspace{-5px} 
			&\!\!\!\!\!\! &  \hspace{-5px}2.999\hspace{-5px}  &  \hspace{-5px}0.089\hspace{-5px}\\
			\!\!\!$\bfR(1,1)$\!\!\! & \!2\!   
			&\!\!\!\!\!\! &  \hspace{-5px}2.001\hspace{-5px}  &  \hspace{-5px}0.100\hspace{-5px} 
			&\!\!\!\!\!\! &  \hspace{-5px}2.002\hspace{-5px}  &  \hspace{-5px}0.100\hspace{-5px} 
			&\!\!\!\!\!\! &  \hspace{-5px}2.001\hspace{-5px}  &  \hspace{-5px}0.059\hspace{-5px} 
			&\!\!\!\!\!\! &  \hspace{-5px}2.003\hspace{-5px}  &  \hspace{-5px}0.058\hspace{-5px}\\
			\!\!\!$\bfR(1,2)$\!\!\! & \!\!\!-1\!  
			&\!\!\!\!\!\! & \hspace{-7px}-1.000\hspace{-5px}  &  \hspace{-5px}0.080\hspace{-5px} 
			&\!\!\!\!\!\! & \hspace{-7px}-0.999\hspace{-5px}  &  \hspace{-5px}0.080\hspace{-5px} 
			&\!\!\!\!\!\! & \hspace{-7px}-0.999\hspace{-5px}  &  \hspace{-5px}0.043\hspace{-5px} 
			&\!\!\!\!\!\! & \hspace{-7px}-0.997\hspace{-5px}  &  \hspace{-5px}0.043\hspace{-5px}\\
			\!\!\!$\bfR(2,2)$\!\!\! & \!1\!   
			&\!\!\!\!\!\! &  \hspace{-5px}1.000\hspace{-5px}  &  \hspace{-5px}0.082\hspace{-5px} 
			&\!\!\!\!\!\! &  \hspace{-5px}1.001\hspace{-5px}  &  \hspace{-5px}0.082\hspace{-5px} 
			&\!\!\!\!\!\! &  \hspace{-5px}0.999\hspace{-5px}  &  \hspace{-5px}0.039\hspace{-5px} 
			&\!\!\!\!\!\! &  \hspace{-5px}1.000\hspace{-5px}  &  \hspace{-10px}0.039\hspace{-5px}
			\\[-11px]
			\\ \cmidrule{1-2} \cmidrule{4-5} \cmidrule{7-8}  \cmidrule{10-11}  \cmidrule{13-14} 
			\\[-10px]
			\!\!\!\text{Time}\!\!\! & - &\hspace{-15px} &  \multicolumn{2}{c}{1$\times$Uw-Nr}  &\hspace{-15px} &  \multicolumn{2}{c}{0.61$\times$Uw-Nr} &\hspace{-15px} &  \multicolumn{2}{c}{5.77$\times$Uw-Nr} &\hspace{-15px} &  \multicolumn{2}{c}{0.80$\times$Uw-Nr} \\
			\toprule
		\end{tabular}
	}
	\caption{Performance MDM Un/semi-weighted non-recursive and recursive identification.}
	\label{tableRec}
    \vspace*{-6mm}
\end{table}
The evolution of the estimates is plotted in Fig.~\ref{recFigtime}.
It can be seen that the MDM Sw-Nr and Sw-Re provide estimates with lower covariance than MDM Uw-Nr and Uw-Re estimates. 
Recursive MDM estimates (i.e., Uw-Re and Sw-Re) are more computationally efficient than non-recursive techniques (i.e., MDM Uw-Nr and Sw-Nr).
\subsection{Assessment of results}
As could be expected, the unweighted MDM identification provides the worst quality results with the smallest computational costs.
On the other hand, the weighted MDM identification offers the highest estimate quality with high costs.
The semi-weighted MDMD identification provides estimate quality comparable to the weighted MDM identification with significantly smaller computational costs.
Note that only weighted MDM identification can additionally provide a covariance matrix of the estimate error that corresponds to the true quality of the estimate. 

The recursive forms are available only for the unweighted and semi-weighted MDM identification.
These forms are suitable not only from the memory storage point of view but also due to their speed.
For example, the recursive form of the semi-weighted MDM identification is smaller in terms of computational costs than the non-recursive unweighted MDM identification.

\section{Concluding Remarks}\label{sec:conclusion}
The paper focused on the noise covariance identification of stochastic linear time-varying state-space models.
A general problem was considered, where the dimension of the measurements and inputs may vary in time.
For such problems, three MDM weighting identification techniques were proposed, which differ in the weighting in the underlying least squares method. For two techniques, their recursive forms were designed. The performance of the proposed techniques has been analyzed using two examples. From the results, it follows that the semi-weighted MDM identification technique, which has not been presented before, provides high-quality estimates with low computational costs, especially in its recursive form. Implementation of all the proposed techniques in \MATLAB is publicly available.

\textbf{Acknowledgment}: \textit{This work was co-funded by the European Union under the project ROBOPROX - Robotics and advanced industrial production  (reg. no. CZ.02.01.01\slash00\slash22\_008\slash0004590).}

\begin{thebibliography}{17}
\providecommand{\natexlab}[1]{#1}
\providecommand{\url}[1]{\texttt{#1}}
\providecommand{\urlprefix}{URL }
\expandafter\ifx\csname urlstyle\endcsname\relax
  \providecommand{\doi}[1]{doi:\discretionary{}{}{}#1}\else
  \providecommand{\doi}{doi:\discretionary{}{}{}\begingroup \urlstyle{rm}\Url}\fi

\bibitem[{Dun\'{i}k et~al.(2018)Dun\'{i}k, Kost, and Straka}]{DuKoSt:18}
Dun\'{i}k, J., Kost, O., and Straka, O. (2018).
\newblock Design of measurement difference autocovariance method for estimation of process and measurement noise covariances.
\newblock \emph{Automatica}, 90, 16--24.

\bibitem[{Dun\'{i}k et~al.(2020)Dun\'{i}k, Kost, Straka, and Blasch}]{DuKoStBl:20}
Dun\'{i}k, J., Kost, O., Straka, O., and Blasch, E. (2020).
\newblock Covariance estimation and gaussianity assessment for state and measurement noise.
\newblock \emph{Journal of Guidance, Control, and Dynamics}, 43(1), 132--139.


\bibitem[{Dun\'{i}k et~al.(2017)Dun\'{i}k, Straka, Kost, and Havl\'{i}k}]{DuStKoHa:17}
Dun\'{i}k, J., Straka, O., Kost, O., and Havl\'{i}k, J. (2017).
\newblock Noise covariance matrices in state-space models: {A} survey and comparison - part {I}.
\newblock \emph{Int. Journal of Adaptive Control and Signal Processing}, 31(11), 1505--1543.

\bibitem[{Huang et~al.(2021)Huang, Zhang, Shi, and Chambers}]{HuZyShCh:21}
Huang, Y., Zhang, Y., Shi, P., and Chambers, J. (2021).
\newblock Variational adaptive Kalman filter with Gaussian-inverse-Wishart mixture distribution.
\newblock \emph{IEEE Transactions on Automatic Control}, 66(4), 1786--1793.

\bibitem[{Kost et~al.(2017)Kost, Dun\'{i}k, and Straka}]{KoDuSt:17}
Kost, O., Dun\'{i}k, J., and Straka, O. (2017).
\newblock Noise covariance matrix estimation in navigation and tracking: Impact of linearisation error.
\newblock In \emph{56th Conference on Decision and Control}. Melbourne, Australia.

\bibitem[{Kost et~al.(2018)Kost, Dun\'{i}k, and Straka}]{KoDuSt:18b}
Kost, O., Dun\'{i}k, J., and Straka, O. (2018).
\newblock Correlated noise characteristics estimation for linear time-varying systems.
\newblock In \emph{57th Conference on Decision and Control}. Miami, FL, USA.

\bibitem[{{Kost} et~al.(2018){Kost}, {Dun\'{i}k}, and {Straka}}]{KoDuSt:18}
{Kost}, O., {Dun\'{i}k}, J., and {Straka}, O. (2018).
\newblock Estimation of noise means and covariance matrices for linear time-varying models.
\newblock \emph{In Proceedings of the 2018 Annual American Control Conference (ACC)}, 265--271.

\bibitem[{Kost et~al.(2018)Kost, Dun\'{i}k, and Straka}]{KoDuSt:18a}
Kost, O., Dun\'{i}k, J., and Straka, O. (2018).
\newblock Noise moment and parameter estimation of state-space model.
\newblock \emph{In Proceedings of 18th IFAC Symposium on System Identification {\nobreak (SYSID)}}, 51(15), 891--896.

\bibitem[{Kost et~al.(2021)Kost, Dun\'{i}k, and Straka}]{KoDuSt:21}
Kost, O., Dun\'{i}k, J., and Straka, O. (2021).
\newblock Identifiability of unique elements of noise covariances in state-space models.
\newblock \emph{Proceedings of 19th IFAC Symposium on System Identification {\nobreak (SYSID)}}, 54(7), 316--321.

\bibitem[{Kost et~al.(2023{\natexlab{a}})Kost, Duník, and Straka}]{KoDuSt:22}
Kost, O., Duník, J., and Straka, O. (2023{\natexlab{a}}).
\newblock Measurement difference method: A universal tool for noise identification.
\newblock \emph{IEEE Transactions on Automatic Control}, 68(3), 1792--1799.

\bibitem[{Kost et~al.(2023{\natexlab{b}})Kost, Duník, Straka, and Daniel}]{KoDuStDa:23}
Kost, O., Duník, J., Straka, O., and Daniel, O. (2023{\natexlab{b}}).
\newblock Identification of gnss measurement error: From time to elevation dependency.
\newblock \emph{IEEE Transactions on Aerospace and Electronic Systems}, 1--12.

\bibitem[{Mehra(1972)}]{Meh:72}
Mehra, R.K. (1972).
\newblock Approaches to adaptive filtering.
\newblock \emph{IEEE Trans. on Automatic Control}, 17(10), 693--698.

\bibitem[{Moghaddamjoo and Kirlin(1993)}]{MoKi:93}
Moghaddamjoo, A.R. and Kirlin, R.L. (1993).
\newblock \emph{Approximate Kalman Filtering}, volume~2 of \emph{Series in Approximations and Decompositions}, chapter Robust Adaptive Kalman Filtering, 65--85.
\newblock World scientific.

\bibitem[{S\"{a}rkk\"{a} and Nummenmaa(2009)}]{SaNu:09}
S\"{a}rkk\"{a}, S. and Nummenmaa, A. (2009).
\newblock Recursive noise adaptive {K}alman filtering by variational {B}ayesian approximations.
\newblock \emph{IEEE Transactions on Automatic Control}, 54(3), 596--600.

\bibitem[{Sch{\"o}n et~al.(2011)Sch{\"o}n, Wills, and Ninness}]{ScWiNi:11}
Sch{\"o}n, T.B., Wills, A., and Ninness, B. (2011).
\newblock System identification of nonlinear state-space models.
\newblock \emph{Automatica}, 47(1), 39--49.

\bibitem[{Solonen et~al.(2014)Solonen, Hakkarainen, Ilin, Abbas, and Bibov}]{SoHaIlAbBi:14}
Solonen, A., Hakkarainen, J., Ilin, A., Abbas, M., and Bibov, A. (2014).
\newblock Estimating model error covariance matrix parameters in extended {K}alman filtering.
\newblock \emph{Nonlinear Processes in Geophysics}, 21(5), 919--927.

\end{thebibliography}

\end{document}